# Investigation of degenerate dual-pump phase sensitive amplifier using multi-wave model

**Weilin Xie,[1,2,*] Ihsan Fsaifes,[1] Tarek Labidi,[1] and Fabien Bretenaker[1]**

[1] *Laboratoire Aimé Cotton, CNRS, Université Paris Sud 11, ENS Cachan, Université Paris Saclay, Campus d'Orsay, 91405 Orsay, France*
[2] *State key Laboratory of Advanced Optical Communication Systems and Networks, Shanghai Jiao Tong University, 800 Dong Chuan Road, Shanghai, 200240, China*
[*]*weilin.xie@u-psud.fr*

**Abstract:** Operation of a degenerate dual-pump phase sensitive amplifier (PSA) is thoroughly numerically investigated using a multi-wave model, taking into account high-order waves associated with undesired four-wave mixing (FWM) processes. More accurate phase-sensitive signal gain spectra are obtained compared to the conventional 3-wave model, leading to the precise optimization of the pump configuration in a dual-pump PSA. The signal gain spectra, as well as the phase sensitivity, are obtained and interpreted by investigating the dominant FWM processes in terms of corresponding phase matching. Moreover, the relation between dispersion slope and gain spectra is revealed, permitting the application-oriented arbitrary tailoring of the gain spectra by manipulating the dispersion profile and pump wavelength allocation.

## 1. Introduction

Phase sensitive amplifiers (PSAs) based on fiber-optic parametric amplifiers (FOPAs) [1], exploiting nonlinear parametric processes in highly nonlinear fiber (HNLF), significantly benefit from intrinsically broadband and noiseless amplification [2-4] and compatibility with current fiber based systems. They thus exhibit attractive prospects in variety of research fields spanning from optical communication, metrology, to signal processing [3-8]. In particular, owing to the essence of ultra-low distortion throughout the entire gain regime, it is of special potential for state-of-the-art microwave photonics (MWP) applications, where ultra-low noise amplification with high linearity and large gain is urgently demanded [6-8]. Compared to single-pump PSA, the dual-pump configuration, capable of providing a broadband flat gain spectrum with less power for each pump, and avoiding the generation of unwanted idler, is of critical interest from the application point of view.

In such dual-pump PSAs, in order to generate a large and flat parametric gain over a broad spectrum, one can use a large wavelength separation between the two pumps that prevents the generation of spurious high-order waves due to multiple four-wave-mixing (FWM) in HNLF. However, phase locking two highly separated pump lasers requires advanced optical injection-locking and optical phase-locking techniques and is still difficult in practice. Moreover, though the use of strong dispersion-slope fiber can suppress the high-order waves, it is unfortunately not favorable for broad bandwidth gain. Additionally, it can be convenient and practical to fit all the waves within the bandwidth of a usual Erbium-doped fiber amplifier (EDFA). Finally, Raman-induced power transfer, which is detrimental to the FWM efficiency [9], is easier to avoid with relatively small pump separations. For all these reasons, it is highly desirable to design a dual-pump PSA with a small pump separation while minimizing the generation of parasitic tones by FWM of the two pumps. Indeed, the existence of these undesired FWM processes associated with high-order waves can affect the phase-sensitive signal gain. To date, the PSA has been both theoretically and experimentally analyzed in depth based on a model describing a single FWM process consisting of 3-wave degenerate FWM (DFWM) or 4-wave non-degenerate FWM (NDFWM) [9-14]. The non-degenerate dual-pump PSA, introducing two additional idlers has been investigated based on the so-called 6-wave model [15-18]. More recently, high-order FWM has been addressed, accounting for sideband-assisted gain extinction ratio enhancement in phase regeneration [19, 20]. However, within the scope of a practical PSA, the thorough investigation and characterization of high-order FWM has been largely overlooked.

In this paper, we focus on the theoretical investigation of a dual-pump degenerate PSA by conducting a multi-wave, more precisely, 7-wave model. Following a 7-wave model rather than the conventional 3-wave model, the impact of the accompanying high-order FWM processes, as well as the relation between signal gain spectra and dispersion, is investigated in terms of gain spectra and power evolution by extensive numerical simulations. Beyond this, we provide physical interpretations of the gain spectra and phase sensitivity, based on the phase mismatch condition with regard to the relevant FWM processes and waves. Thanks to this physical interpretation, we can predict which processes limit the efficiency of the PSA. In particular, the phase sensitive gain spectra can be precisely tailored and manipulated, thus enabling application-oriented optimization of various PSAs. This is particularly interesting for MWP links, where small pump separations can be sufficient owing to the limited bandwidth of the amplified signals.

## 2. Multi-wave Model

The concept of the dual-pump degenerate PSA as well as a preliminary experiment result are shown in Fig. 1. First, Fig. 1(a) shows what happens when co-polarized signal $S_0$ and pumps $P_1$ and $P_2$ are launched in a $L = 1011$-m-long HNLF (OFS standard HNLF) with a nonlinear coefficient $\gamma = 11.3$ W$^{-1}$.km$^{-1}$. The pump-pump separation is set to 40 GHz, with a central wavelength equal to $\lambda_{ZDW} = 1547.5$ nm, corresponding to the zero-dispersion wavelength of the fiber. The total input power of the two pumps is equal to 23.5 dBm. At the output of the HNLF [see Fig. 1(a)], many high-order waves are generated by cascaded FWM of the two pumps and the pumps and signal. This leads to a significant deterioration of the signal gain compared to the value estimated from the 3-wave model. The spectral broadening of the high-order waves is due to the phase modulation of the pumps used to suppress stimulated Brillouin scattering (SBS). In order to investigate more accurately such a situation, we use a 7-wave model, as depicted in Fig. 1(b). Beyond the initially launched signal and pumps $S_0$, $P_1$, and $P_2$, we introduce the waves labeled 3 and 4 mainly generated by FWM of the signal with pumps 1 and 2, respectively, and the waves labeled 5 and 6 mainly generated by FWM of the pumps. $A_j$ and $\omega_j$ ($j = 0..6$) represent the complex field amplitudes and frequencies of the waves. $\Delta\lambda_{PP}$ is the pump-pump wavelength separation while the wavelength offset $\delta\lambda_{ofs} = \lambda_0 - \lambda_{ZDW}$ corresponds to the deviation of the signal wave $\lambda_0$ with respect to the zero dispersion wavelength $\lambda_{ZDW}$ of the fiber. In the following, $A_3$, $A_4$ and $A_5$, $A_6$ are also called high-order signals and pumps, respectively.

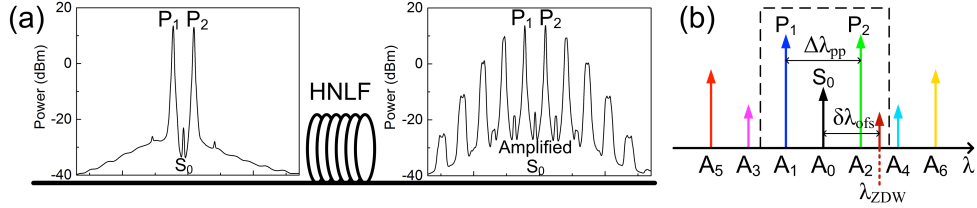

Fig. 1. (a) Experimental result of dual-pump degenerate PSA: considerable high-order waves are generated by underlying high-order FWM process. (b) Labeling of waves, separation, and offset for 7-wave model. The three waves in the dashed-line rectangle are those considered in the fundamental 3-wave model.

The field evolution of the seven co-polarized waves co-propagating in the $z$ direction along the fiber with length $L$, attenuation coefficient $\alpha$, nonlinear coefficient $\gamma$, and dispersion slope $D_\lambda$ is governed by a set of seven complex coupled equations [9]. For the sake of clarity, we reproduce here only one of these equations, e.g. for signal field $A_0$:

$$\begin{aligned}\frac{dA_0}{dz} = -\frac{\alpha}{2}A_0 + i\gamma\Bigg\{&\left[|A_0|^2 + 2\sum_{i=0,i\neq 0}^{6}|A_i|^2\right]A_0 + A_1^2 A_3^* e^{-i\Delta\beta_{0311}z} + A_2^2 A_4^* e^{-i\Delta\beta_{0422}z} \\ &+ 2A_1 A_2 A_0^* e^{-i\Delta\beta_{0012}z} + 2A_1 A_6 A_4^* e^{-i\Delta\beta_{0416}z} + 2A_1 A_3 A_5^* e^{-i\Delta\beta_{0513}z} + 2A_1 A_4 A_2^* e^{-i\Delta\beta_{0214}z} \\ &+ 2A_2 A_5 A_3^* e^{-i\Delta\beta_{0325}z} + 2A_2 A_3 A_1^* e^{i\Delta\beta_{2301}z} + 2A_2 A_4 A_6^* e^{i\Delta\beta_{2406}z} + 2A_4 A_5 A_1^* e^{i\Delta\beta_{4501}z} \\ &+ 2A_5 A_6 A_0^* e^{i\Delta\beta_{5600}z} + 2A_3 A_4 A_0^* e^{i\Delta\beta_{3400}z} + 2A_3 A_6 A_2^* e^{i\Delta\beta_{3602}z}\Bigg\}\end{aligned} \qquad (1)$$

where the superscript * holds for complex conjugate. The term containing $\alpha$ accounts for the fiber attenuation, the two square modulus terms inside the square bracket are responsible for nonlinear phase shifts due to self-phase modulation (SPM) and cross-phase modulation (XPM), respectively, and the other terms correspond to the energy transfers between the interacting waves due to FWM processes. All the relevant waves are equally spaced in frequency on the opposite sides of the signal wave $\omega_0$ in a symmetrical fashion due to energy conservation $\omega_m + \omega_n = \omega_k + \omega_l$ and $2\omega_m = \omega_k + \omega_l$ ($\omega_m + \omega_n = 2\omega_k$) representing the NDFWM and DFWM processes respectively. If we consider one of the NDFWM processes, for which $A_k$ and $A_l$ play the roles of the pumps and $A_m$ and $A_n$ those of the signal and idler, the corresponding linear phase mismatch can be written as

$$\Delta\beta_{mnkl} = \beta_m + \beta_n - \beta_k - \beta_l , \qquad (2)$$

where $\beta_i$ is the wavevector of $A_i$ calculated at its frequency $\omega_i$. This is done by expanding $\beta$ to $4^{th}$ order in Taylor power series around the signal frequency $\omega_0$. The high-order derivatives of the propagation constant are deduced from the fiber dispersion coefficients. Equation (2) is also valid for DFWM cases with either $\beta_m = \beta_n$ or $\beta_k = \beta_l$.

This set of complex coupled equations is quite general in the sense that it includes effects such as depletion, high-order dispersion, and nonlinear phase shifts. For the 7-wave model, all together 13 NDFWM and 9 DFWM processes are taken into account. The extension to more interacting waves could improve the accuracy to some extent, especially for small values of $\Delta\lambda_{PP}$, though at the expense of a much more complicated set of coupled equations due to the contribution of many more involved FWM processes. This would make any physical interpretation of the results almost impossible. The 7-wave model, which exhibits the similar tendency as models involving more waves and offers sufficient estimation accuracy with sustainable complexity, is thus adopted. By solving the set of complex coupled differential equations simultaneously in a numerical manner, one can obtain the field evolution of each wave along the fiber.

## 3. Signal Gain Spectra using 7-wave Model

The signal gain spectra are numerically obtained using the following HNLF parameters, which we will keep throughout the paper: $L = 1011$ m, $\gamma = 11.3$ $W^{-1}.km^{-1}$, $\alpha = 0.9$ dB/km, and dispersion slope $D_\lambda = 0.017$ $ps.km^{-1}.nm^{-2}$. For $\delta\lambda_{ofs} = 0$, Fig. 2 compares these results with those from the 3-wave model. The incident wave powers are 100 mW for each pump and 1 µW for the signal: we are thus in the small signal regime. All calculations are performed for a zero relative phase between the launched signal and pump waves.

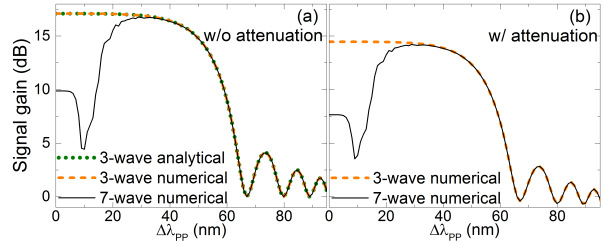

Fig. 2. Signal gain spectra calculated when $\delta\lambda_{ofs} = 0$ using 3- and 7-wave models (a) analytically and numerically without depletion; (b) numerically with depletion, respectively. The fiber parameters are given in the text.

The 3-wave analytical model [9] [see the dashed line in Fig. 2(a)], is valid only if the pumps remain undepleted. Depletion can be included only by solving the 3-wave coupled equations numerically [see the dashed line in Fig. 2(b)]. Neglecting the fiber attenuation, both 3-wave models give the same prediction, while the 7-wave model indicates serious gain distortion in small $\Delta\lambda_{PP}$ region [see Fig 2(a)]. This is a strong indication in favor of the usefulness of the 7-wave model. Not surprisingly, if attenuation is taken into account in both

3- and 7-wave numerical solutions, the gain values become smaller compared to the previous cases. From Fig. 2, it is clear that the 7-wave model exhibits improved accuracy for estimating and investigating the practical gain spectra, especially in the small $\Delta\lambda_{PP}$ region where the 3-wave model is clearly invalid.

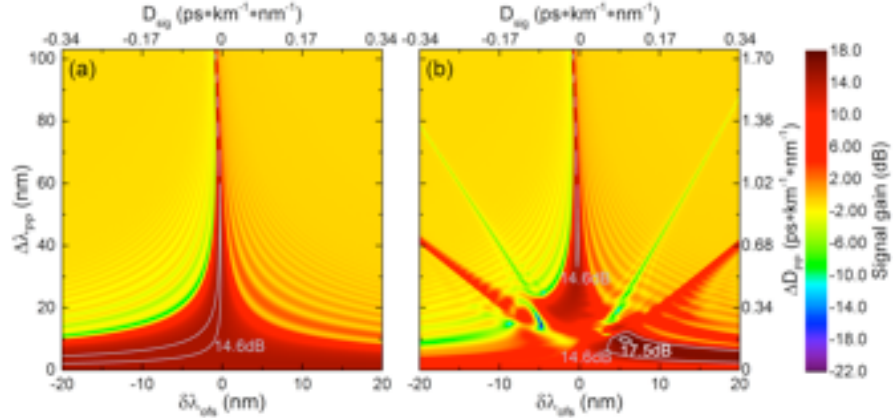

Fig. 3. Signal gain spectra vs. signal wavelength offset $\delta\lambda_{ofs}$ with respect to the zero dispersion wavelength $\lambda_{ZDW}$ and vs. pump-pump separation $\Delta\lambda_{PP}$ for the (a) 3- and (b) 7-wave models. The right axis corresponds to the dispersion difference $\Delta D_{PP}$ between the two pumps while the upper axis corresponds to the dispersion $D_{sig}$ for the signal.

By varying $\delta\lambda_{ofs}$ and $\Delta\lambda_{PP}$ simultaneously while keeping the relative phase between the input signal and pumps and the input powers fixed, we generate the signal gain heatmaps reproduced in Fig. 3. Figures 3(a) and 3(b) are obtained using the 3-wave and 7-wave models, respectively. When the signal wave is in the vicinity of $\lambda_{ZDW}$, though significant gain deterioration is observed when the pumps are closely located, the gain bandwidth is maximized and tends to that of the 3-wave model when $\Delta\lambda_{PP}$ is larger than 30.0 nm. However, compared to the 3-wave model, the peak gain in the 7-wave model is achieved when the signal wavelength is larger than $\lambda_{ZDW}$ by about 5 nm. The predicted gain maximum is then even higher (17.5 dB) than the one (14.6 dB) predicted by the 3-wave model. When the signal is located in the anomalous dispersion regime ($\delta\lambda_{ofs} > 0$), the decrease of the gain with $\Delta\lambda_{PP}$ follows a similar tendency in the two models. Conversely, when the signal lies in the normal dispersion regime ($\delta\lambda_{ofs} < 0$), the 7-wave model predicts serious gain distortions at small $\Delta\lambda_{PP}$, including peaks and dips, which are absent from the 3-wave model predictions. Again, in the range 20 nm $\leq \Delta\lambda_{PP} \leq$ 100 nm, the two models exhibit similar behaviors. One striking feature of the 7-wave model with respect to the 3-wave one is that the gain peak vanishes and the gain bandwidth decreases rapidly when the signal is moving further towards the normal dispersion regime, leaving only smaller gain peaks and dips around small $\Delta\lambda_{PP}$ region, especially between – 10 nm and 0 deviation from $\lambda_{ZDW}$. The actinomorphic gain peaks and dips can be attributed to some phase matching situations at certain wavelength configurations, and will be investigated in the next section.

## 4. Physical Interpretation of Gain Spectra

According to the phase matching essence of FWM [9, 21], the process efficiency is governed by the effective phase mismatch $\kappa_{mnkl}$ of the considered FWM process occurring between waves $m$, $n$, $k$, and $l$

$$\kappa_{mnkl} = \Delta\beta_{mnkl} + \gamma P_{mnkl} ,  \qquad (3)$$

where $\Delta\beta_{mnkl}$ is the linear phase mismatch term (see eq. 2). $\gamma P_{mnkl}$ is the nonlinear phase mismatch term, which depends on the powers of the involved waves, through the relation

$$\gamma P_{mnkl} = \gamma\left(P_k + P_l - P_m - P_n\right), \qquad (4)$$

where $P_i$ is the power of wave $i$. Compared to the fundamental 3-wave model where only $\kappa_{0012} = \Delta\beta_{0012} + \gamma P_{0012} = (2\beta_0 - \beta_1 - \beta_2) + \gamma(P_1 + P_2 - 2P_0)$, associated with the DFWM of the initial three waves, is relevant, the 7-wave model involves all the 22 FWM processes occurring simultaneously along the fiber, leading to much more complicated situations. For the sake of clarity, the physical interpretation of the 7-wave model will be given by observing when these processes are phase matched, and thus expected to play a significant role. We do this by plotting the relevant $\kappa_{mnkl}$ associated with the FWM process we want to consider and that we suspect to lead to a significant energy transfer between $A_k$, $A_l$ and $A_m$, $A_n$. The nonlinear part of these $\kappa_{mnkl}$ is calculated with the values of the powers $P_{mnkl}$ obtained at the end of the fiber. In these plots, we multiply $\kappa_{mnkl}$ by the fiber length: this leads to a phase mismatch expressed in units of an angle, and we expect the considered FWM process to be efficient only when the absolute value of this angle is small compared to $\pi$.

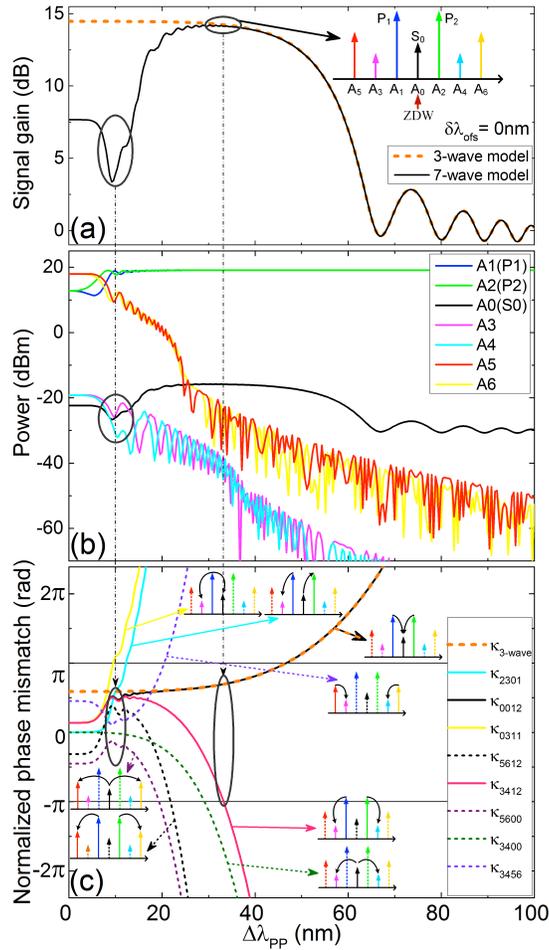

Fig. 4. Case where $\delta\lambda_{ofs} = 0$ nm. (a) Signal gain spectra for the 3- and 7-wave models versus pump-pump wavelength separation $\Delta\lambda_{PP}$; the inset shows the wavelengths configuration relatively to the zero-dispersion wavelength ZDW. (b) Corresponding evolution of the output powers of the seven waves. (c) Normalized phase mismatch angles of the relevant FWM processes. The input phases of the signal and pumps are chosen to correspond to the maximum gain of the PSA in the ordinary 3-wave regime. The regions featured by ellipses are discussed in the text, and the corresponding dominant processes are given in the insets.

*4.1 Zero dispersion region*

Let us start by considering the situation where $\delta\lambda_{ofs} = 0$, for which $\lambda_{ZDW}$ is at the center of all the waves, as illustrated in the inset of Fig. 4(a). Then, the different FWM processes involving waves located in a symmetrical manner with respect to $\lambda_{ZDW}$ can achieve perfect phase matching at some values of $\Delta\lambda_{PP}$. Figure 4(a) represents the corresponding evolution of the signal gain versus $\Delta\lambda_{PP}$, while Fig. 4(b) reproduces the corresponding output powers of the 7 waves versus $\Delta\lambda_{PP}$. Fig. 4(b) shows that the high-order signals and pumps (waves labeled 3, 4, 5, and 6), although they emerge only from the combination of high-order FWM processes, can exhibit significant output powers, even stronger than the incident signal and pumps, respectively. This happens for small values of $\Delta\lambda_{PP}$ ($\Delta\lambda_{PP} < 20$ nm). This is explained by the values of the phase mismatch coefficients of the FWM processes that generate these high-order waves, as shown in Fig. 4(c). Indeed, in this region, the phase mismatch coefficients $\kappa_{0012}$, $\kappa_{5612}$, $\kappa_{3412}$, $\kappa_{3400}$, $\kappa_{5600}$, $\kappa_{3456}$, remain within the $\pm\pi$ range for which the corresponding processes are efficient. As suggested by these $\kappa$'s, besides the fundamental 3-wave phase mismatch $\kappa_{0012}$, which governs the energy transfer between pumps and signal, the energy is directed towards the high-order signals and pumps power from the input pumps and even from the input signal. This leads to the observed drastic pump depletion and the severe signal gain distortion. This is particularly striking at $\Delta\lambda_{PP} = 10$ nm, where $\kappa_{3400}$ and $\kappa_{5600}$ are close to 0, leading to almost perfect phase matching for the corresponding processes, explaining the remarkable signal gain dip at such value of $\Delta\lambda_{PP}$. Beyond 20 nm separation, all the spurious processes have $\kappa$ values outside the $\pm\pi$ range and vanish, leaving only the fundamental process, namely $\kappa_{0012}$, within the $\pm\pi$ range: the gain predicted by the 3-wave model is then retrieved. It is worth noticing that, $\kappa_{3400}$ and $\kappa_{5600}$ become phase mismatched twice and three times quicker than $\kappa_{0012}$, respectively, due to the fact that these processes involve high-order signals and pumps with twice or three times larger frequency separations than the incident waves. This makes their linear phase mismatches much more sensitive to the increase of $\Delta\lambda_{PP}$. Similar behavior can also be found for $\kappa_{3456}$ and $\kappa_{5612}$. Consequently, the signal gain completely retrieves the values predicted by the 3-wave model for $\Delta\lambda_{PP} \geq 30$ nm, where only $\kappa_{0012}$ dominates over all the other processes, which are completely phase mismatched.

*4.2 Normal dispersion region*

Let us now turn to the case where $\delta\lambda_{ofs} = -10$ nm, as shown in Fig. 5(a-c). Despite some gain peaks, the gain spectrum predicted by the 7-wave model is completely different from the one derived from the 3-wave model. For small values of $\Delta\lambda_{PP}$, owing to the intricate interplay of many processes whose values of $\kappa$ are within the $\pm\pi$ range, one can hardly distinguish the dominant ones. The main FWM process associated with the phase mismatch $\kappa_{0012}$ fades rapidly when $\Delta\lambda_{PP}$ increases because its phase mismatch exits the $\pm\pi$ range as soon as $\Delta\lambda_{PP} \geq 12$ nm. The signal gain becomes then extremely small. When $\Delta\lambda_{PP}$ reaches about 15 nm, the process governed by $\kappa_{0624}$ becomes dominant, as indicated in Fig. 5(c). This can be easily understood as $A_0$ and $A_2$ on the one hand and $A_4$ and $A_6$ on the other hand are then almost symmetrical with respect to $\lambda_{ZDW}$, as shown in the inset of Fig. 5(a), thus approaching perfect phase matching. Through this process, $A_0$ and $A_6$ gain energy from $A_2$ and $A_2$, leading to the fact that $A_6$ becomes stronger than $A_5$ in the neighboring $\Delta\lambda_{PP}$ region. In the same region, the gain and power of $A_0$ start to increase a bit. Interestingly, $A_4$ exhibits some power losses for certain values of $\Delta\lambda_{PP}$, and for some others maintains a non-negligible level thanks to the FWM associated with $\kappa_{4426}$. When we further increase $\Delta\lambda_{PP}$, $\kappa_{1604}$, $\kappa_{0422}$, and $\kappa_{1622}$ play an important role around $\Delta\lambda_{PP} = 20$ nm, leading to a significant power transfer from $A_0$, $A_2$, and $A_4$ to the other involved waves. This happens because $A_2$ is nearly located at $\lambda_{ZDW}$ and all the FWM processes that are symmetric with respect to it are experiencing perfect phase matching. Thus it turns out that the powers of $A_6$, $A_4$, and $A_1$ are more significant than those of $A_5$, $A_3$, and $A_2$, respectively, in the vicinity of such values of $\Delta\lambda_{PP}$. Specifically, even the pump

$A_1$ gets amplified owing to these processes. Beyond 30 nm separation, the signal gain almost vanishes and no significant depletion is observed, leaving only some tiny ripples in large $\Delta\lambda_{PP}$ regions. However, around $\Delta\lambda_{PP} = 40$ nm, one can observe a significant and narrow gain dip attributed to the phase matching of $\kappa_{1402}$, $\kappa_{3614}$, and $\kappa_{3602}$, as indicated in Fig. 5(c). The processes pump power out of $A_0$. In summary, throughout this normal dispersion region, one can hardly achieve an optimum pump configuration for signal amplification. This makes this regime unsuitable for applications.

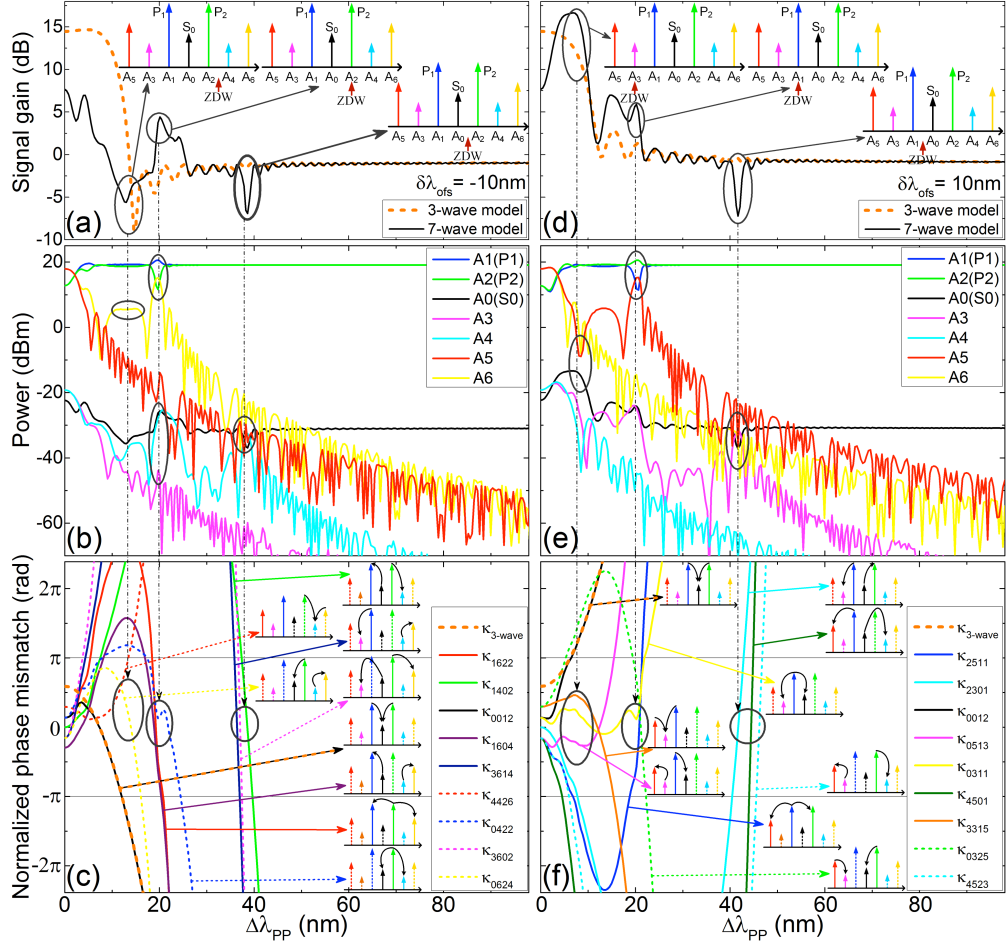

Fig. 5. (a-c) Same as Fig. 4 for $\delta\lambda_{ofs} = -10$ nm (signal in normal dispersion region). (d-f) Same as Fig. 4 for $\delta\lambda_{ofs} = 10$ nm (signal in anomalous dispersion region).

*4.3 Anomalous dispersion region*

We finally turn to the opposite detuning $\delta\lambda_{ofs} = 10$ nm. The signal is now located in the anomalous dispersion regime and a dramatic gain hump is observed for $\Delta\lambda_{PP} \approx 5$ nm [see Fig. 5(d)], with much higher gain than predicted by the 3-wave model. This is mainly attributed to the fact that $A_0$ undergoes amplification not only thanks to the fundamental FWM process governed by $\kappa_{0012}$, which indeed remains between $-\pi$ and $\pi$, but also thanks to two other phase matched FWM processes corresponding to $\kappa_{0513}$ and $\kappa_{0311}$ [see Fig. 5(f)]. Such a sideband-assisted gain enhancement has remained largely unexplored previously and will be further investigated below. As $\Delta\lambda_{PP}$ increases to larger separation regions, this gain peak decreases rapidly. A second but smaller gain peak is found at $\Delta\lambda_{PP}$ = 15 nm. Although the process governed by $\kappa_{0513}$ is no longer phase matched, this peak is explained by the process associated with $\kappa_{0311}$, which remains pretty well phase matched. In the vicinity of $\Delta\lambda_{PP}$ = 15 nm, $A_5$, $A_3$, $A_1$, and $A_0$ are almost symmetric with respect of $\lambda_{ZDW}$, explaining the nearly perfect phase matching of $\kappa_{0513}$, $\kappa_{3315}$ and $\kappa_{0311}$. Therefore, it turns out that the power of $A_5$ and $A_3$ is larger than that of $A_6$ and $A_4$ owing to the energy transfers induced by the corresponding FWM processes. At about $\Delta\lambda_{PP}$ = 20 nm, a situation similar to the one we met for $\delta\lambda_{ofs}$ = –10 nm occurs. Indeed, since $A_0$ is located nearby $\lambda_{ZDW}$, the waves positioned at symmetric positions with respect to $A_0$ can experience phase matched FWM. The FWM processes governed by $\kappa_{0325}$ and $\kappa_{2511}$, become predominant in addition to the main one governed by $\kappa_{0012}$. This gives rise to the third gain peak in Fig. 5(d). Since $\kappa_{0012}$ is already far away from the $\pm\pi$ range, the two secondary gain peaks are smaller than the main one occurring at $\Delta\lambda_{PP} \approx 5$ nm. In the adjacent region for which 12 nm $\leq \Delta\lambda_{PP} \leq$ 20 nm, the power evolutions of $A_5$, $A_3$, and $A_2$, and of $A_6$, $A_4$, $A_1$, exhibit opposite evolutions with respect to the preceding case for which $\delta\lambda_{ofs}$ = –10 nm [compare Figs. 5(b) and 5(e)]. This is consistent with the fact that the phase mismatches for these processes have opposite values in the two cases $\delta\lambda_{ofs}$ = 10 nm and $\delta\lambda_{ofs}$ = –10 nm . Moreover, similarly to the normal dispersion regime case, a narrow gain dip is observed around $\Delta\lambda_{PP}$ = 40 nm, owing to the fast evolution of $\kappa_{2301}$, $\kappa_{4523}$, and $\kappa_{4501}$ around zero. This also leads to the power increase of $A_3$. Finally, compared to the case where the signal is located in the normal dispersion regime, and in certain regions within the anomalous dispersion regime, the $\gamma P_{mnkl}$ term can cancel the $\Delta\beta_{mnkl}$ in $\kappa_{mnkl}$, leading to better phase matching for the signal wave and subsequently to significant signal gain peaks for specific values of the pump-pump wavelength separation.

*4.4 Discussion*

In all three cases considered above, the signal gain becomes negligibly small beyond 40 nm pump-pump separation, a region in which the PSA becomes unusable for applications due to the large phase mismatch of the principal FWM process. The power evolution and signal gain spectra, which we have just seen to be governed by the phase matching conditions, can be directly extended to other pump frequency allocations. Remarkably, although the $\kappa$'s are simply evaluated at the fiber output, rather than integrated along the fiber, the simple criterion consisting in looking whether the phase mismatch angle is within the $\pm\pi$ range or not has been shown to be relevant to evaluate whether a given FWM process is dominant or negligible. This shows that, contrary to what could have been expected, one can build some intuition of what happens in such complicated multi-wave nonlinear problems. It is worth noting that the directions of the energy flows indicated in the insets of Figs. 4 and 5 are directly deduced from the power evolutions of the corresponding waves. They could be also obtained by looking at the relative phase of each process along the fiber. However, this is quite complicated and well beyond the scope of the present paper.

As mentioned above, the phase matching conditions, which determine the signal gain spectrum, depend on the wave powers, pump spectral positions and thus dispersion properties of the fiber according to eq. (3). To investigate the dispersion dependence of the gain spectra, we change the dispersion slope $D_\lambda$ of the HNLF with fixed initial wave powers. The resultant gain spectra scale inversely proportional to $D_\lambda$, as illustrated in Fig. 6. Thus $\Delta\lambda_{PP}$ and $\delta\lambda_{ofs}$ can be directly transposed in terms of dispersion profile, as illustrated by the top and right axes of Figs. 3(a) and 3(b). For these axes, $D_{sig}$ is the dispersion at the signal wavelength and $\Delta D_{PP}$ is the dispersion difference between two pumps. The gain spectra can be subsequently normalized by the dispersion profile of the HNLF. When combined with the 7-wave model, the precise tailoring of the gain spectra by manipulating the dispersion profile permits a full optimization of the dual-pump PSA gain from the application point of view, such as the low distortion and low noise amplification on a single carrier in MWP links.

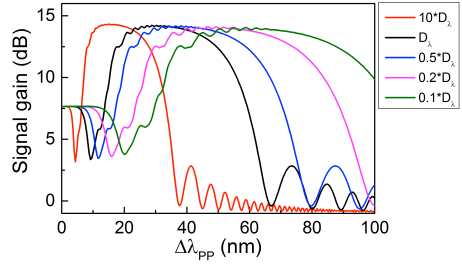

Fig. 6. Signal gain spectra calculated for different factors multiplying the dispersion slope $D_\lambda$ when $\delta\lambda_{ofs} = 0$. The initial wave powers remain the same for all the values $D_\lambda$.

*4.5 Phase-sensitivity of the extra gain predicted by the 7-wave model*

For the sake of thoroughly characterizing and investigating the phase sensitivity of the signal gain peak obtained in anomalous dispersion regime, we plot the gain versus relative phase of the waves at various values of $\Delta\lambda_{PP}$ with $\delta\lambda_{ofs} = +10$ nm, as shown in Fig. 7.

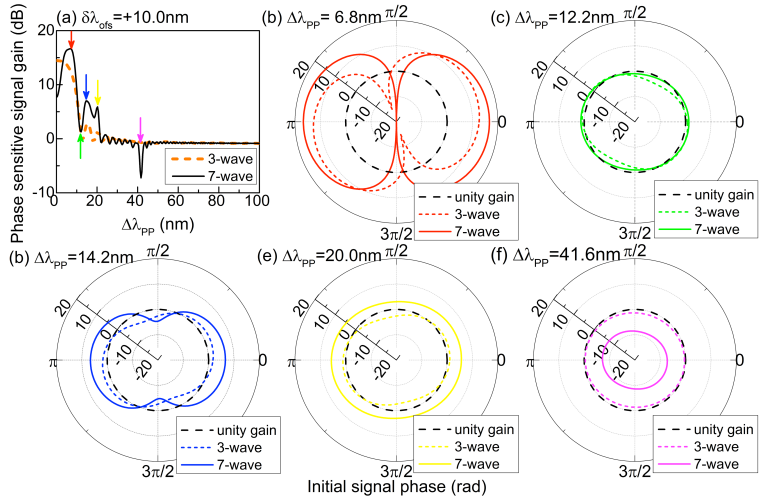

Fig. 7. (a) Same gain versus pump-pump separation profile as in Fig. 5(d), obtained for $\delta\lambda_{ofs} = +10$ nm. The vertical arrows point at the values of $\Delta\lambda_{PP}$ corresponding to (b-f). (b-f) Gain (in dB) polar plot versus relative phase between the signal and the pumps for the 3-wave (dashed line) and 7-wave (full line models) for (b) $\Delta\lambda_{PP} = 6.8$ nm (first gain peak in (a)), (c) $\Delta\lambda_{PP} = 12.2$ nm (first gain dip in (a)), (d) $\Delta\lambda_{PP} = 14.2$ nm (second gain peak in (a)), and (e) $\Delta\lambda_{PP} = 20.0$ nm (third gain peak in (a)), (f) $\Delta\lambda_{PP} = 41.6$ nm (second gain dip in (a)), .

This figure shows how the gain depends on the input relative phase between the three waves for different values of the pump-pump separation $\Delta\lambda_{PP}$, indicated by the arrows in the gain spectrum of Fig. 7(a). A very interesting feature can be noticed in Fig. 7(b), which corresponds to the situation where the 7-wave model predicts more gain than the 3-wave model. Quite remarkably, this large gain is indeed shown to be phase sensitive, with an extinction ratio larger than 36 dB, larger than predicted with the 3-wave model. It is worth noticing also that the maximum and minimum gains are slightly phase shifted compared to the 3-wave model. A similar behavior is observed at the second gain peak position, as shown in Fig. 7(d). Conversely, the third gain peak is subject to less phase-sensitivity as a result of the corresponding large value of $\kappa_{0012}$, which shows that the fundamental gain process is no longer active. Additionally, since the dominant processes around the third gain peak are governed by $\kappa_{0325}$ and $\kappa_{0311}$, and thus involve not only the initial three waves but also waves emerging from high-order FWM processes, we notice a strong degradation in the degree of phase-sensitivity of this gain and of its extinction ratio.

As shown in Fig. 7(c), the phase sensitive gain around the first gain dip retrieves a similar tendency as in the 3-wave model, which well agrees with the spectra in Fig. 7(a) in the vicinity of this dip. As the gain varies around 0 dB, it exhibits weak phase-sensitivity. Similarly, though the phase-sensitivity is also observed at the narrow second gain dip with de-amplification as indicated in Fig. 7(f), the dominant FWM processes which decreases the signal power involves high-order waves $A_3$, $A_4$, and $A_5$ besides the fundamental ones, thus impairing the phase-sensitivity at this dip.

## 5. Conclusion

In conclusion, high-order waves originating from the high-order FWM processes have been shown to be properly described in the framework of a 7-wave model. Numerical integration of this model has led to accurate signal gain spectra for the degenerate dual-pump PSA. It turns out that a PSA with an appropriate choice of wavelengths can achieve even higher signal gain than the one expected from the conventional 3-wave model, thanks to the extra gain provided by high-order FWM processes associated with high-order waves. The gain spectra also revealed the regions where efficient gain can be obtained in different wave configurations, as well as some non-efficient configurations that should be avoided from the application point of view.

The physical interpretation of the complicated gain spectra has been elaborated by further investigation of the dominant FWM processes in terms of the corresponding phase matching conditions. In addition, the phase sensitivity of the signal gain has been analyzed. The gain has been shown to be more or less phase sensitive in several PSA configurations, depending on whether the dominant FWM processes involve not only the fundamental 3-wave DFWM but also other higher-order FWM processes or not. Moreover, the gain spectra are shown to be scalable along with the dispersion of the fiber, permitting an arbitrary tailoring of these spectra by manipulating the fiber dispersion profile. With the proposed 7-wave model, application-oriented arbitrary gain spectra can be achieved together with a PSA configuration, which is optimized from a practical point of view. Especially for MWP links applications, where a large peak gain with low distortion and low noise is preferred rather than a broadband flat gain spectrum, the multi-wave model can easily select the most efficient configuration in view of maximizing the gain peak.

## Acknowledgements

This work is partially supported by the French Agence Nationale de la Recherche under contract No. ANR-12-BS03-001-01, Thales Research & Technology, and the Chinese Government Scholarship (CSC) under grant 201406230161.